\documentclass[trackchanges]{aastex701}
\usepackage{amsmath}
\usepackage{subcaption}
\usepackage{graphicx}
\usepackage{import}
\newcommand{\q}[2]{#1_{#2}}
\newcommand{\qz}[2]{\tilde{#1}_{#2}}

\begin{document}

\title{MAS-CCD: New technique for measuring low-level charge content based on the multiple amplifier architecture}

\author[0009-0006-6711-6050]{Miqueas E. Gamero}
\affiliation{Instituto de Inv. en Ing. Eléctrica (IIIE), DIEC, Universidad Nacional del Sur (UNS)-CONICET, Bahía Blanca, Argentina}
\affiliation{Fermi National Accelerator Laboratory, Batavia, IL, USA.}
\email{miqueas.gamero@uns.edu.ar}

\author[]{Guillermo Fernández Moroni}
\affiliation{Fermi National Accelerator Laboratory, Batavia, IL, USA.}
\affiliation{Department of Astronomy and Astrophysics, University of Chicago, Chicago IL, USA.}
\email{gfmoroni@fnal.gov}

\author[]{Fernando Chierchie}
\affiliation{Instituto de Inv. en Ing. Eléctrica (IIIE), DIEC, Universidad Nacional del Sur (UNS)-CONICET, Bahía Blanca, Argentina}
\email{}

\author[]{Agustin J. Lapi}
\affiliation{Fermi National Accelerator Laboratory, Batavia, IL, USA.}
\affiliation{University of Chicago, Chicago IL, USA.}
\email{}

\author[]{Blas J. Irigoyen Gimenez}
\affiliation{Instituto de Inv. en Ing. Eléctrica (IIIE), DIEC, Universidad Nacional del Sur (UNS)-CONICET, Bahía Blanca, Argentina}
\affiliation{Fermi National Accelerator Laboratory, Batavia, IL, USA.}
\email{}

\author[]{Juan Estrada}
\affiliation{Fermi National Accelerator Laboratory, Batavia, IL, USA.}
\affiliation{Department of Astronomy and Astrophysics, University of Chicago, Chicago IL, USA.}
\affiliation{Brookhaven National Laboratory, Upton, NY, USA}
\email{}

\author[]{Javier Tiffenberg}
\affiliation{Fermi National Accelerator Laboratory, Batavia, IL, USA.}
\email{}

\begin{abstract}

Low-noise detectors are a key technology for the next generation of astronomical instruments aimed at spectroscopy of faint objects and the search for exoplanets. In this context, the multiple-amplifier sensing charge-coupled device (MAS-CCD) emerges as a promising technology for future scientific instruments. A critical parameter affecting the performance of these devices is spurious charge, produced by the clocking of the gates. Its measurement is typically challenging with existing methods. In practice, the optimization of this parameter often relies on empirical procedures that require significant time and careful consideration of the trade-off with full-well capacity.

In this work, we present a new technique to estimate spurious charge based on covariance analysis of the output amplifiers of the MAS-CCD, which measures the same charge packet in different amplifiers at different times. The method enables fast and precise measurements of spurious charge under operating conditions where conventional approaches are difficult to apply. We develop the theoretical framework of the method and validate the model through simulations. The results demonstrate the feasibility of this approach and suggest that it could serve as a basis for reliable large-scale characterization of sensor performance.

\end{abstract}


\section{Introduction}

\begin{figure}[h!]
    \centering

    \begin{subfigure}{0.69\textwidth}
        \centering
        \includegraphics[width=\linewidth]{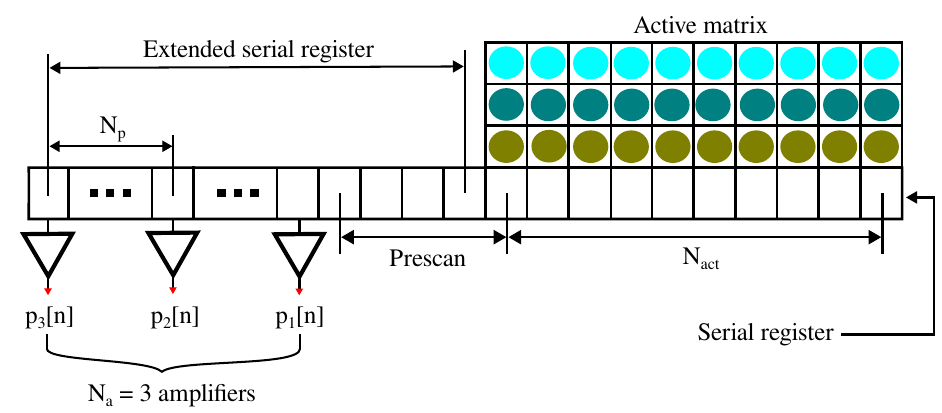}
        \caption{Sketch of the MAS-CCDs architecture.}
        \label{fig:masccd-sketch}
    \end{subfigure}
    \hfill
    
    \begin{subfigure}{0.69\textwidth}
        \centering
        \includegraphics[width=\linewidth]{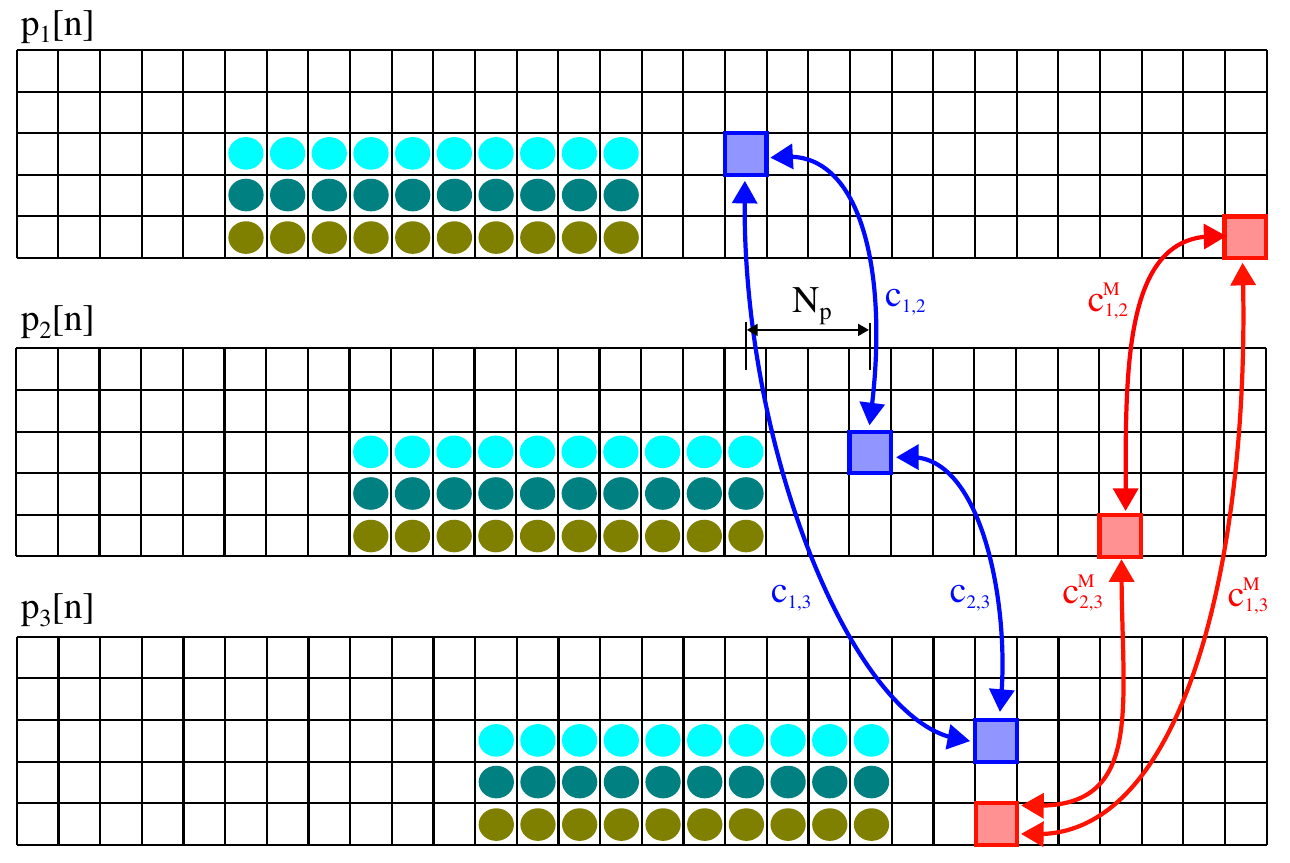}
        \caption{Graphical representation of the pixel offsets and random variables definitions for the calculation of the covariance values using pixels from different amplifiers when they are in charge phase (blue) and in noise phase (red).}
        \label{fig:images-definitions}
    \end{subfigure}
    \caption{Sketch of simplified MAS CCD output stage and its output images.}
    \label{fig:masccd-full}
\end{figure}

The Multi Amplifier Sensing Charge Coupled Device (MAS CCD or MAS-CCD) offers a powerful architecture to reduce readout noise without increasing the sensor's readout time. The MAS-CCD is based on many floating gate output stages on the same serial register \cite{Holland_SDW23, Blas2024}. Figure \ref{fig:masccd-sketch} shows a diagram of its architecture. As the pixels are transferred through the serial register, the different output stages measure the pixel's content. Then, the information from the different amplifiers for the same active area pixel is combined into a final pixel value. Since the samples from the different amplifiers are independent, the final pixel value has readout noise smaller by a factor equal to the square root of the number of amplifiers. Its ability to reduce readout noise with a negligible impact on the readout time has encouraged its use in the next generation of astronomical instruments \cite{Lapi2024MAS16, Lin_2024}. Two examples are its use in a new terrestrial instrument to make an extensive survey of the spectrum of distant galaxies for universe expansion studies \cite{Spec-S5:2025uom}, and its use to measure the spectrum of exoplanet atmospheres to find Earth-like planets from space-based observatories \cite{mas-ccd-NASA}.

Once the readout noise of the sensor is reduced, other intrinsic sources of noise define the new uncertainty limitation in the pixel charge measurements. One of the dominant sources in CCDs is the spurious charge (SC), also referred to as clock-induced charge (CIC), produced by the clocking of the gates to move the charge in the pixels of the array \cite{janesick2001scientific, cababie_2022}. In particular, for astronomical applications, since the swings of the clocks are large to be able to transfer large amounts of charge (several thousand carriers), the SC could have a significant impact on the measurement error of the charge collected by the pixels \cite{10.1111/j.1365-2966.2010.17675.x, Lin_2024}. In particular, the SC produced in the serial register has been measured as the most significant contribution to errors \cite{Cervantes-Vergara_2023, Villalpando_2024}.

Measuring the SC using traditional techniques is challenging due to the convolution of the Poisson distribution of the SC with the Gaussian distribution of the readout noise. Since the Gaussian component includes an arbitrary offset introduced by the amplifier electronics, it is difficult to disentangle the mean of the Poisson-distributed SC from the mean of the Gaussian noise. An alternative approach is to attempt to infer the presence of SC through the overall width, or standard deviation, of the distribution. However, this method is also imprecise because, for low mean values of SC, the total width is dominated by the readout noise, making the SC contribution difficult to isolate and quantify accurately.

These challenges can be addressed by leveraging the intrinsic features of the MAS-CCD architecture. Since the charge signal is correlated across all output channels—being derived from the same pixel—while the noise contributions from the amplifiers are uncorrelated, it is possible to separate the two random process using correlation analysis. By computing the correlation function among amplifier outputs, one can isolate the signal-related (correlated) components from the noise (uncorrelated) components. This approach provides a promising method to efficiently decouple the contribution of SC from readout noise, enabling more accurate characterization of SC effects even when their mean levels are low.

Section \ref{sec:operation} describes the main architectural components of the MAS-CCD and introduces mathematical models for the intrinsic error sources in the device. Section \ref{sec:estimation-two-pixels} takes an initial approach to estimating the SC from the covariance of two pixels carrying the same information but originating in different channels. Section \ref{sec:variance-estimator} presents the variance of the proposed estimator. Section \ref{sec:general-framework} explains the general framework for estimating the signal with an arbitrary number of channels and presents an optimal way of combining their information to improve the estimation. Section \ref{sec:correlated-noise} describes how to handle a possible source of correlated noise among channels. For an extension of this method to dark-current estimation, refer to Section \ref{sec:dark-current}. Finally, Section \ref{sec:implementation} presents the implementation of the technique along with simulations that validate its functioning. Appendix \ref{ap:derivations} includes the full mathematical derivations used across the different sections.

\section{MAS-CCD operation and intrinsic noise source modeling}\label{sec:operation}

A simplified layout of a 3-amplifier MAS-CCD is presented in Fig. \ref{fig:masccd-sketch}. During the readout operation, each row of the active matrix is transferred to the serial register, where it is sequentially read out through the output amplifiers. The prescan area is used to facilitate the transport of charge packets to the output stage. For simplicity, in the present analysis, these few pixels are not considered. Once a row has been read out, the next row is transferred to the serial register and undergoes the same readout process. Thus, although the image is displayed as a 2D matrix, the readout scheme can be interpreted as a 1D stream of data that is later rearranged into a 2D image. Since each amplifier records data independently, the readout system generates distinct images corresponding to each amplifier. Figure \ref{fig:images-definitions} depicts three distinct output images obtained from each amplifier. The spatial shift between pixels with the same information is due to the extra transfers, $N_p$, required for the charge packets to reach the next amplifier. Although the amplifiers are read out simultaneously, each one samples a different pixel at a given time. Consequently, the same pixel in the active region is measured by different amplifiers at different times. The resulting pixel information for amplifier $i$, where $1 \leq i \leq N_a$, can be defined as $p_{i}[n]$, where $n$ is the index of the pixel in the output image.

The pixel value is typically modeled as a sum of two terms,
\begin{equation}
    p_{i}[n] = r_{i}[n] + q_i[n] = w_i[n] + m_i[n] + q_i[n],
\end{equation}
where $q_i[n]$ is the charge measured by the $i$-th amplifier at time $n$ and $r_i[n]$ is the readout noise of the measurement, which can further be decomposed into two terms: an independent term, $w_i[n]$, given by the noise of each transistor, and a correlated noise source, $m_i[n]$, originating from common control signals, electromagnetic interference, or grounding loop problems. To simplify the mathematical expressions from now on, it is assumed that all these signals are in units of equivalent collected carriers, i.e., they have been calibrated using the measured gain of the system. 

As seen in Figure \ref{fig:masccd-sketch}, the same charge packet is measured by two amplifiers, $i$ and $j$, assuming $i<j$, with a offset equal to $(j-i)N_p$. Therefore, a positive correlation is expected in the pixel values from different amplifiers due to the charge information if a proper shift of $(j-i)N_p$ is applied to the index of the output images. This alignment is referred to as the ``charge phase,'' illustrated by the blue pixels and definitions in Figure \ref{fig:images-definitions}. This correlation results from the charge packet measured, either SC or active area charge.

To establish the mathematical framework for evaluating this correlation, we define the random variable $\q{P}{1}=p_{1}[n_0]$ for a pixel of the first amplifier at an arbitrary pixel index $n_0$. Then, for the following amplifiers, the respective random variable in charge phase with the first amplifier is defined as $\q{P}{i}=p_{i}[n_0+(i-1)N_p]$. Then, it can be express as
\begin{equation}
    \q{P}{i}=\q{R}{i}+\q{Q}{i}=\q{W}{i}+\q{M}{i}+\q{Q}{i},
\end{equation}
where $\q{W}{i} = w_i[n_0+(i-1)N_p]$, $\q{M}{i}=m_i[n_0+(i-1)N_p]$, and $\q{Q}{i}=q_i[n_0+(i-1)N_p]$. For the mathematical framework, the following statistical properties of the defined random variables are assumed:
\begin{itemize}
     \item $\q{W}{i}\sim G(\mu_i,\sigma_i)$ follows a Gaussian distribution. The offset of this distribution is an analog offset added by the video output electronic chain. This variable is modeled to be statistically independent of the same variables of other amplifiers, i.e., $E[\q{W}{i}\q{W}{j}]=0$ for $j\neq i$.
     
    \item $\q{M}{i}$ is the correlated noise between channels. Since this noise arises from imperfections in the system, it is challenging to establish a probability distribution. However, previous experience has shown that it can be well-modeled as a stationary process and as a joint stationary process for the contributions from different amplifiers. 
    
    \item $\q{Q}{i}$ is the charge signal to be estimated. Several models could contribute here, but it is possible to separate the contributions to the charge produced in the active region, which remains constant during its transit in the serial register, from the charge generated in the serial register, which may change during the readout process. Two examples of the former are the intrinsic leakage current of the device and other background photons collected by the pixels in the active region during the sensor's exposure time, as well as the SC generated by the vertical clocking, also known as vertical SC. Similar processes affect the pixels in the serial register. Since the effective exposure time of the pixels in the serial register is typically much smaller than their time in the vertical array, the leakage current contribution here is relatively small. However, the SC contribution in the serial register cannot be typically neglected. These processes, in both regions, can be well modeled using a Poisson distribution:
    \begin{equation}
     \q{Q}{i}\sim P(\lambda_v + \lambda_i),
    \end{equation}
    where $\lambda_v$ accounts for the mean of the charge collected by the pixels while being in the active matrix, and $\lambda_i$ accounts for the amount of SC generated in the serial register until it reaches the sense node of amplifier $i$. From now on, unless explicitly mentioned, the focus will be on the SC produced in the serial register, also known as horizontal SC. 
\end{itemize}

The horizontal SC is typically estimated from the overscan region of the image, where there is no contribution from charge in the active region. In this region, the pixels have a mean charge of
\begin{equation}\label{eq:lambda and sc}
    \lambda_i=\mathrm{SC}(N_{\text{act}}+(i-1)N_p),
\end{equation}
where the SC is in units of electrons per pixel per transfer, $N_{\text{act}}+(i-1)N_p$ is the number of transfers a pixel in the overscan undergoes to reach amplifier $i$ and $N_{\text{act}}$ is the number of pixels in the active matrix.

An important remark is that the variables $\q{W}{i}$, $\q{M}{i}$, and $\q{Q}{i}$ are independent of each other, so their covariances are equal to zero. We also define an auxiliary variable
\begin{equation}\label{eq:extra-variableQij}
    \q{Q}{i,j}\sim P(\lambda_{i,j}), \quad \lambda_{i,j}= \mathrm{SC} (j-i)N_p , \text{ with } i<j,
\end{equation}
which models the charge that was generated between amplifiers $i$ and $j$. Then,
\begin{equation}
     \q{Q}{j} = \q{Q}{i,j} + \q{Q}{i}.
\end{equation}
Here, $\q{Q}{i,j}$ and $\q{Q}{i}$ are independent. Therefore,
\begin{equation}
    E\left[\q{Q}{i},\q{Q}{i,j}\right]=0.
    \label{eq: indep of charge}
\end{equation}

During the standard analysis of the output images, the global offset of each channel is subtracted from the image. This offset is typically estimated as the mean of pixels from the overscan region. After this step, the pixel model is:
\begin{equation}
    \qz{P}{i}=\q{P}{i}-E[\q{P}{i}]=\qz{W}{i} + \qz{M}{i} + \qz{Q}{i},
\end{equation}
where $\qz{W}{i} = \q{W}{i}-\q{\mu}{i}$, $\qz{M}{i} = \q{M}{i}-E[\q{M}{i}]$, and $\qz{Q}{i} = \q{Q}{i}-\lambda{i}$. This is the main reason why the mean of the Poisson distribution of the SC cannot be directly estimated from its mean value, as this mean value is lost when subtracting the offset added by the electronic chain to the output video signal.

\section{Estimation of the charge signal using the covariance of the pixels from two channels} \label{sec:estimation-two-pixels}

To estimate the $\mathrm{SC}$ parameter using the covariance of pixels in charge phase, it is convenient to define the product of pixels from channels $i$ and $j$ as
\begin{equation}\label{eq:Cu}
C_{i,j}=\qz{P}{i}\qz{P}{j},
\end{equation}
where $i < j$. The covariance can be expressed as 
\begin{equation}
\begin{split}
    c_{i,j} &=E[C_{i,j}]=E[\qz{P}{i}\qz{P}{j}] = E[(\qz{W}{i} + \qz{M}{i} + \qz{Q}{i})(\qz{W}{j} + \qz{M}{j} + \qz{Q}{j})] = E[\qz{M}{i}\qz{M}{j}] + E[\qz{Q}{i}\qz{Q}{j}].
\end{split}
\end{equation}
The simplified two-term formula follows from the independence among variables. Further decomposition of the expression yields:
\begin{equation}\label{eq:expectation}
\begin{split}
    c_{i,j} = E[\qz{Q}{i}(\qz{Q}{i} + \qz{Q}{i,j})] + E[\qz{M}{i}\qz{M}{j}] = E[\qz{Q}{i}^2] + E[\qz{M}{i}\qz{M}{j}] = \lambda_{i} + E[\qz{M}{i}\qz{M}{j}],
\end{split}
\end{equation}
where $E[\qz{Q}{i}^2] = E[(Q_i - \lambda_i)^2] = \lambda_i$. The results from Eq. (\ref{eq: indep of charge}) have been used to simplify the first term. Eq. (\ref{eq:expectation}) represents an important result: the estimator can be biased if there is a correlation among channels in the system. To simplify the analysis in the following sections, we assume that there is no correlated noise affecting the channels, or mathematically, $E[\qz{M}{i}\qz{M}{j}]=0$. This noise source will be addressed in a later section. Using Eq. (\ref{eq:lambda and sc}), the covariance can be rewritten as
\begin{equation}
c_{i,j}=\lambda_{i}=\mathrm{SC}(N_{\text{act}}+(i-1)N_p).
\label{eq: covij}
\end{equation}
This leads to the important remark that the covariance is only defined by the charge produced before the $i$-th amplifier when $i<j$. Finally, the SC can be calculated as the covariance of the pixels from different amplifiers in charge phase as
\begin{equation}\label{eq:sc}
    \mathrm{SC} = \frac{c_{i,j}}{(N_{\text{act}}+(i-1)N_p)} = \frac{\lambda_i}{(N_{\text{act}}+(i-1)N_p)}.
\end{equation}

\subsection{Charge estimation on actual images}

The covariance, $c_{i,j}$, is estimated from a large number of pixels across the output images. The estimation begins by computing the mean-subtracted pixel values within the region of interest used to estimate the charge contribution, mathematically:
\begin{equation}
    \qz{p}{i}[n] = p_{i}[n]-\hat{p}_{i}, \text{ and } \qz{p}{j}[n] = p_{j}[n]-\hat{p}_{j},
    \label{eq: signal minus mean}
\end{equation}
where
\begin{equation}
\hat{p}_{i} = \sum_{s=1}^{N_{\text{pix}}}\frac{p_{i}[s]}{N_{\text{pix}}}, \quad \hat{p}_{j} = \sum_{s=1}^{N_{\text{pix}}}\frac{p_{j}[s + \tau]}{N_{\text{pix}}},
\end{equation}
$\tau$ is the pixel offset between the channels and $N_{\text{pix}}$ is the total number of pixels available in the region of interest. Note that, for the two channels to be in charge phase, the condition $\tau = \tau_{i,j} = (j-i)N_p$ must be satisfied. A consistent mathematical formulation for the covariance as a function of the offset $\tau$, is
\begin{equation}
\text{Cov}(\qz{p}{i}, \qz{p}{j})[\tau] = \sum_{s=1}^{N_{\text{pix}}}\frac{(p_{i}[s]-\hat{p}_{i})(p_{j}[s + \tau]-\hat{p}_j)}{N_{\text{pix}}-1}.
\label{eq: cov estimator}
\end{equation}
Finally, the estimation in charge phase can be expressed as:
\begin{equation}
\hat{c}_{i,j} = \text{Cov}(\qz{p}{i}, \qz{p}{j})\big\rvert_{\tau=(j-i)N_p}.
\label{eq:cov-estimator-cp}
\end{equation}
This analysis, has been focused solely on the SC produced in the horizontal register. In this case, the region of pixels must be in the overscan region of the images.

\section{Variance of the estimator}
\label{sec:variance-estimator}

The variance of the proposed estimator gives the baseline for the convergence power of the methodology when applied to real images. It is still assumed that the correlated noise is not present to simplify the equations. In Appendix \ref{ap:derivations}, the derivation of the estimator variance is given in full detail. Here, however, a simpler case that arises from the complete derivation in the aforementioned Appendix is presented. The simplified variance expression is
\begin{equation}
\label{eq:estimator-variance}
\begin{split}
\text{Var}[C_{i,j}]&=\text{Var}[\qz{P}{i}\qz{P}{j}]=\text{E}[(\qz{P}{i}\qz{P}{j} - c_{i,j})^{2}] = \sigma_i^2\sigma_j^2 + \sigma_i^2\lambda_{j}+\sigma_j^2\lambda_{i}+\text{E}[\qz{Q}{i}^4]-\lambda_{i}^2 + \lambda_{i}\lambda_{i,j},
\end{split}
\end{equation}
where 
\begin{equation}
\text{E}[\qz{Q}{i}^4]=\text{E}[(Q_i-\lambda_i ) ^4]=3\lambda_{i}^2  +\lambda_{i}
\end{equation}
is the fourth central moment of $\q{Q}{i}$. Eq. (\ref{eq:estimator-variance}) shows the penalization of using a second-order estimator, $E[\qz{P}{i}\qz{P}{j}]$, to estimate the mean charge value, since its variance has fourth-power terms of the noise in the first term and of the charge amount in the fourth term.

\section{General framework for the estimation of the low-level charge in the pixels} \label{sec:general-framework}

The MAS-CCD sensors have multiple output amplifiers. Currently available devices have up to 16 output channels. Therefore, an extended framework allowing for the use of the covariance information from multiple pairs of channels is required. However, the total number of available covariance pairs, $N_c = N_a(N_a-1)/2$, can be impractical to work with for $N_a = 16$ or larger. It is then convenient to simplify the notation by defining a unique index $u$ ($1 \le u \le N_c$) that unambiguously identifies every pairwise combination:
\begin{equation}
u=f(i,j)=j-i + \sum_{z=1}^{i-1}(N_a-z), \quad 1\leq i<j \leq N_a.
\label{u index}
\end{equation} 
Thus, it is appropriate to handle the available covariance information in matrix form,
\begin{equation}
	\mathbf{C}^T = \text{E}[\q{C}{1}\dots \q{C}{u} \dots \q{C}{N_c}]=[c_1\dots c_u \dots c_{N_c}].
\label{eq: matrix covariance}
\end{equation} 
Introducing a weight vector to address the different contributions from each covariance,
\begin{equation}
	\mathbf{A}^T = [a_1 \dots a_{N_c}],
\end{equation}
allows to estimate the SC information as
\begin{equation}
\mathrm{SC} = \mathbf{A}^T\mathbf{C}.
\label{eq:sc from vectors}
\end{equation}
Assuming there is no correlated noise among channels, $\mathbf{C}$ can be expressed, using Eq. (\ref{eq: covij}) and (\ref{u index}), as
\begin{equation}
\begin{aligned} \label{eq:decomposition}
\mathbf{C}^T
 &= [\lambda_{1}, \dots, \lambda_{1},
     \lambda_{2}, \dots, \lambda_{2},
     \dots,
     \lambda_{N_a-2}, \lambda_{N_a-2},
     \lambda_{N_a-1}] \\
 &= \mathrm{SC}\bigl[
      N_{\text{act}}, \dots, N_{\text{act}},
      N_{\text{act}} + N_p, \dots, N_{\text{act}} + N_p, \\
 &\qquad\qquad
      \dots,
      N_{\text{act}} + (N_a-3)N_p,
      N_{\text{act}} + (N_a-3)N_p,
      N_{\text{act}} + (N_a-2)N_p
    \bigr] \\
 &= \mathrm{SC}\,\mathbf{D}^T,
\end{aligned}
\end{equation}
where $\mathbf{D}$ is the distance vector that represents the number of charge transfers each pixel undergoes before reaching the $i$-th amplifier. Replacing the result of Eq. (\ref{eq:decomposition}) in Eq. (\ref{eq:sc from vectors}) yields
\begin{equation}
    1 = \mathbf{A}^T\mathbf{D},
    \label{eq:restriction}
\end{equation}
which defines the condition on the weights under which the combination of the available covariances yields the desired estimate. However, this equation does not provide a closed-form solution to find the weights. One of the solutions to the equation is given by
\begin{equation}
    a_u = \frac{1}{N_{\text{act}}+(i-1)N_p} \text{ with } u=f(i,j), 1\leq i<j \leq N_a.
    \label{eq:simple-weights}
\end{equation}
This scales each covariance by the additional contribution of pixel transfers needed to reach the next amplifier. Nevertheless, it is not statistically the optimal solution, as there could exist a correlation between the different covariances. However, as will be presented in the following section, the optimal solution is very close to the one from Eq. (\ref{eq:simple-weights}) for most cases of practical interest. 

\subsection{Optimal weights for the spurious charge estimator}
\label{sec: optimal weights}

The variance of the estimator introduced in Eq. (\ref{eq:sc from vectors}) is used as the merit function for the optimization, subject to the restriction in Eq. (\ref{eq:restriction}):
\begin{equation}\label{eq:min_var}
\begin{aligned}
& \underset{\mathbf{A}}{\text{minimize}} & &  \text{Var}[\mathbf{A}^T \mathbf{C}] \\
& \text{subject to} & & 1 = \mathbf{A}^T\mathbf{D}. \\
\end{aligned}
\end{equation}
It is favorable to represent the variance in its matrix form,
\begin{equation}\label{eq:solution optimization}
	\text{Var}[\mathbf{A}^T \mathbf{C}] = \mathbf{A}^T \mathbf{\Sigma} \mathbf{A},
\end{equation}
where $\mathbf{\Sigma}$, the covariance matrix, is
\begin{equation}\label{eq:Sigma matrix}
	\mathbf{\Sigma} = 
	\begin{bmatrix}
	s_{1,1} & s_{1,2} & \dots & s_{1,N_c}\\
	s_{2,1} & s_{2,2}& \dots &s_{2,N_c}\\
	\vdots & \vdots & \ddots & \vdots\\
	s_{N_c,1} & \dots & \dots& s_{N_c,N_c}\\
	\end{bmatrix}.
\end{equation}
Each element, $s_{u,v}$, represents the covariance between the products of pixels $C_u$ and $C_v$; 
\begin{equation}
\begin{aligned}
    s_{u,v} &=\mathrm{Cov}(C_u,C_v)=\mathrm{Cov}(C_{i,j},C_{k,l}) = \mathrm{E}[(\qz{P}{i}\qz{P}{j}-\mathrm{E}[\qz{P}{i}\qz{P}{j}])(\qz{P}{k}\qz{P}{l}-\mathrm{E}[\qz{P}{k}\qz{P}{l}])], 
\end{aligned}
\label{eq: matri components}
\end{equation}
where distinct, unambiguous index redefinitions, $u = f(i, j), \ (i<j)$ and $v = f(k, l), \ (k<l)$, were introduced for convenience. To be able to compute the covariance matrix, it is necessary to have a closed-form expression for its elements. Below, the full form of $s_{u,v}$ is presented:
\begin{equation}
\label{eq:cov-expression}
\begin{aligned}
s_{u,v} &=
  \delta(h_1-h_3)\delta(h_2-h_4)\sigma_{h_1}^2\sigma_{h_2}^2 \\
&\quad + \delta(h_1-h_3)\sigma_{h_1}^2\lambda_{h_2}
      + \delta(h_1-h_4)\sigma_{h_1}^2\lambda_{h_2} \\
&\quad + \delta(h_2-h_3)\sigma_{h_2}^2\lambda_{h_1}
      + \delta(h_2-h_4)\sigma_{h_2}^2\lambda_{h_1} \\
&\quad - \lambda_{h_1}\lambda_{h_2}
      + \mathrm{E}\!\left[\qz{Q}{h_1}\qz{Q}{h_2}\qz{Q}{h_3}\qz{Q}{h_4}\right].
\end{aligned}
\end{equation}
Two important remarks derive from Eq. (\ref{eq:cov-expression}). First, $\delta(x)$ is a discrete delta function, which serves to compact the final expression. Second, a re-indexing is necessary for generality: $h_1$, $h_2$, $h_3$, and $h_4$ map to $i$, $j$, $k$, and $l$ under the condition $h_1 \leq h_2 \leq h_3 \leq h_4$. For a detailed derivation of Eq. (\ref{eq:cov-expression}), refer to Appendix~\ref{ap:derivations}. 

To minimize Eq. (\ref{eq:solution optimization}), the objective function is defined using Lagrangian multipliers for the constraints in Eq. (\ref{eq:min_var}):
\begin{equation}
     \mathbf{J} =\mathbf{A}^T\mathbf{\Sigma} \mathbf{A} - \beta \left(\mathbf{A}^T \mathbf{D}-1 \right).
\end{equation}
Its solution can be found by differentiating the objective function with respect to the unknown weights, $\mathbf{A}$, and the Lagrangian multiplier, $\beta$,
\begin{equation}\label{eq:derivate objective}
\begin{cases}
     \frac{\partial \mathbf{J}}{\partial \mathbf{A}} = 2\mathbf{A}^T\Sigma - \beta \mathbf{D}^T = \mathbf{0}^T,\\
     \frac{\partial \mathbf{J}}{\partial \beta} = \mathbf{A}^T \mathbf{D} - 1 = 0.
\end{cases}
\end{equation}
The expression for the weights, $\mathbf{A}^T$, can be found using the first derivative equation:
\begin{equation}\label{eq:weiths equation}
\mathbf{A}^T = \frac{\beta {\mathbf{D}^T} \mathbf{\Sigma}^{-1}}{2}.
\end{equation}
To determine the Lagrangian multiplier, $\beta$, the result in Eq. (\ref{eq:weiths equation}) is replaced in the second derivative:
\begin{equation}\label{eq:beta}
    \beta = \frac{2}{\mathbf{D}^T\mathbf{\Sigma}^{-1}\mathbf{D}}.
\end{equation}
Finally, Eq. (\ref{eq:weiths equation}) and Eq. (\ref{eq:beta}) are combined to yield the final expression for the optimal weights:
\begin{equation}
        \mathbf{A}^T = \frac{\mathbf{D}^T \mathbf{\Sigma}^{-1}}{\mathbf{D}^T \mathbf{\Sigma}^{-1} \mathbf{D}}.
        \label{eq:optimal-solution}
\end{equation}

It is of particular interest to discuss Eq. (\ref{eq:cov-expression}). In a real application, low SC levels are desired, that is, the mean SC should be much lower than the standard deviation of the readout noise ($\lambda_{i}<<\sigma_j$ for all $1\le i, \ j \le N_a$). Under these circumstances, the first term is much larger than the sum of the rest of the terms. In this scenario, the covariance matrix in Eq. (\ref{eq:optimal-solution}) behaves like a diagonal matrix, yielding a result similar to the simplified solution in Eq. (\ref{eq:simple-weights}). Another interesting aspect of Eq. (\ref{eq:optimal-solution}) is that it provides a tool for calculating optimal weights in scenarios where correlated noise between channels can be a significant contributor to the covariance estimator uncertainty. For this purpose, the elements of the covariance matrix can be estimated by implementing the expectation in Eq. (\ref{eq: matri components}) using the pixels of the actual output images of the system. It will be presented in the following sections how it is possible to reduce the contribution of correlated noise in the special case where this extra noise in the channels is jointly stationary.

\section{Handling the correlated noise between channels} \label{sec:correlated-noise}

In the presence of correlated noise affecting multiple channels, various approaches are available to minimize the error in calculating the charge correlation between channels. One approach is to eliminate this contribution from the pixel information of the output images before the covariance calculation. For example, the authors in \cite{Lapi2024MAS16, Lin_2025} demonstrate how to estimate correlated noise in one channel using pixel information from the other channels. Other authors in \cite{10.1093/mnras/stv2410, Fernandez_Moroni_2020} use the time correlation of the noise to correct the error in the pixels using other pixels in the same image. The correlated noise is estimated and subtracted from the corresponding channels. 

In this article, we propose an alternative approach oriented toward the presented covariance technique for estimating the charge, assuming that the noise in the affected channels is jointly stationary. The red variables depicted in Fig. \ref{fig:images-definitions} illustrate how a new set of covariances for pixels from different amplifiers can be defined in the ``noise phase''. As evident from the drawing, in this case the covariance is computed using an offset equal in magnitude and opposite in sign to that applied in the charge phase. 

As a starting point for the mathematical modeling, a single pixel of the first amplifier is considered, $\qz{p}{1}[n_0]$, at an arbitrary pixel $n_0$. The values of the mean-subtracted pixels, as defined in Eq. (\ref{eq: signal minus mean}), are used across this section. For the $i$-th amplifier, the pixel in the noise phase relative to the first amplifier is defined as $\qz{p}{i}[n_0-(i-1)N_p]$. This definition can be extended to calculate the covariance in noise phase between two amplifiers, $i$ and $j$, as
\begin{equation*}
\begin{aligned}
\mathrm{Cov}(\qz{p}{i}[n_i],\qz{p}{j}[n_j])
 &= E[\qz{p}{i}[n_i]\qz{p}{j}[n_j]] \\
 &= E[(\qz{r}{i}[n_i]+\qz{q}{i}[n_i]+\qz{m}{i}[n_i])
      (\qz{r}{j}[n_j]+\qz{q}{j}[n_j]+\qz{m}{j}[n_j])].
\end{aligned}
\end{equation*}
where $n_i = n_0 - (i-1)N_p$, $n_j = n_0 - (j-1)N_p$, and $i < j$. Since $\qz{r}{i}$ and $\qz{r}{j}$ have zero expectation and are independent of each other and of the remaining signals, these terms can be removed. As the correlated noise in the pixel measurement is independent of the collected charge, the covariance may be written as:
\begin{equation}
\label{eq:noise_covariance}
\begin{aligned}
\mathrm{Cov}(\qz{p}{i}[n_i],\qz{p}{j}[n_j])
 &= E[(\qz{q}{i}[n_i]+\qz{m}{i}[n_i])(\qz{q}{j}[n_j]+\qz{m}{j}[n_j])] \\
 &= E[\qz{q}{i}[n_i]\qz{q}{j}[n_j]]
  + E[\qz{q}{i}[n_i]\qz{m}{j}[n_j]] \\
 &\quad
  + E[\qz{m}{i}[n_i]\qz{q}{j}[n_j]]
  + E[\qz{m}{i}[n_i]\qz{m}{j}[n_j]] \\
 &= E[\qz{q}{i}[n_i]]E[\qz{q}{j}[n_j]]
  + E[\qz{q}{i}[n_i]]E[\qz{m}{j}[n_j]] \\
 &\quad
  + E[\qz{m}{i}[n_i]]E[\qz{q}{j}[n_j]]
  + E[\qz{m}{i}[n_i]\qz{m}{j}[n_j]] \\
 &= E[\qz{m}{i}[n_i]\qz{m}{j}[n_j]] .
\end{aligned}
\end{equation}

For this analysis, it is assumed that the correlated noise between different amplifiers is a jointly wide-sense stationary stochastic process. In this case, the covariance between two noise signals depends only on the difference between their indices and not on their absolute values. Using this property, the expected value can be expressed as
\begin{equation}
\label{eq:result_equation_corr_noise}
\begin{aligned}
E[\qz{m}{i}[n_i]\qz{m}{j}[n_j]]
 &= \mathrm{Cov}(\qz{m}{i}[n_i],\qz{m}{j}[n_j]) \\
 &= \mathrm{Cov}_{(\qz{m}{i},\qz{m}{j})}[n_i-n_j] \\
 &= \mathrm{Cov}_{(\qz{m}{i},\qz{m}{j})}[n_j-n_i] \\
 &= \mathrm{Cov}(\qz{m}{i}[n_j],\qz{m}{j}[n_i]) \\
 &= E[\qz{m}{i}[n_j]\qz{m}{j}[n_i]] .
\end{aligned}
\end{equation}

Analogous to the approach used in Eq. (\ref{eq:noise_covariance}), the last expectation can also be calculated directly from the pixel values:
\begin{equation}
\begin{aligned}
	\text{Cov}(\qz{p}{i}[n_j],\qz{p}{j}[n_i]) = E[\qz{m}{i}[n_j]\qz{m}{j}[n_i]].
\end{aligned}
\label{eq:result_equation_corr_noise 2}
\end{equation}
Eqs. (\ref{eq:result_equation_corr_noise}) and (\ref{eq:result_equation_corr_noise 2}) show that the correlated noise contribution added to the covariance in charge phase is the same as that added to the covariance calculated in noise phase. In summary, the error due to the correlated noise affecting the calculation of charge information can be estimated as
\begin{equation}\label{eq: cov noise phase}
\begin{split}
c^M_{i,j} = E[C^M_{i,j}] = \text{Cov}(\qz{p}{i}[n_i],\qz{p}{j}[n_j]) = E[\qz{m}{i}[n_i]\qz{m}{j}[n_j]] = E[\qz{M}{i}\qz{M}{j}],
\end{split}
\end{equation}
which is referred to as the covariance in noise phase. This contribution can be estimated from the output data and used to correct the value of the covariance in charge phase. Therefore, reformulating Eq. (\ref{eq:expectation}):
\begin{equation}\label{eq:cov correctoin}
    c_{i,j} - c^M_{i,j} = \lambda_{i} + E[\qz{M}{i}\qz{M}{j}] - c^M_{i,j} = \lambda_{i},
\end{equation}
where $c^M_{i,j}$ is estimated from the output images using the same expressions in equations (\ref{eq: signal minus mean}) through (\ref{eq: cov estimator}), but using the appropriate index offset of $\tau^M_{i,j} = -\tau_{i,j}=(i-j)N_p$ for the $i$ and $j$ channels, $i<j$.

\section{Extension to dark current estimation} \label{sec:dark-current}

The same calculations can be used to estimate small charge contributions to the pixels from the active area. In this case, the pixel charge information is not influenced by the number of transfers in the horizontal register and is therefore independent of the amplifier index. Under this condition, the covariance can be stated as
\begin{equation}
c_{i,j}=\lambda_{v} \text{ for any } 1\leq i,\ j \leq N_a.
\label{eq: covij 2}
\end{equation}
This result simplifies the vector of all possible covariance expectations from Eq. (\ref{eq:decomposition}) to
\begin{equation*}
	\mathbf{C}^T = \lambda_{v}[1, \dots, 1],
\end{equation*} 
whose size is $N_c$. The average value of $\mathbf{C}^T$ yields an estimate of the mean charge in the given region of interest, $\lambda_{v}$. To account for statistical correlation between channels and possible external noise sources, the mathematical framework in Section \ref{sec: optimal weights} can be used, with an updated definition of the distance vector, $\mathbf{D}=[1, \dots, 1]$. The optimal weights are then calculated using the same weight expression as in Eq. (\ref{eq:optimal-solution}). Finally, the charge per pixel is derived as
\begin{equation}
\lambda_{v} = \mathbf{A}^T\mathbf{C}.
\label{eq:lambda vertical from vectors}
\end{equation}

\section{Implementation of the proposed technique} \label{sec:implementation}

\subsection{Simulated images and covariance estimation}
\label{sec:simulations}

To evaluate the performance of the proposed technique, a set of simulations was conducted. These simulated images include different noise conditions and SC levels; only electronic readout noise and horizontal SC were considered. The simulated device dimensions are: $N_a = 16$ output stages, $N_p = 15$ pixels between amplifiers, $N_{\text{act}} = 512$ columns, and $1024$ rows in the active region. However, the final simulated images have a size of $6000$ rows by $900$ columns in order to mimic the region of interest over which covariances are estimated in real images.

\begin{figure}[h!]
  \centering
  \begin{subfigure}{0.45\textwidth}
    \centering
    \includegraphics[width=\linewidth]{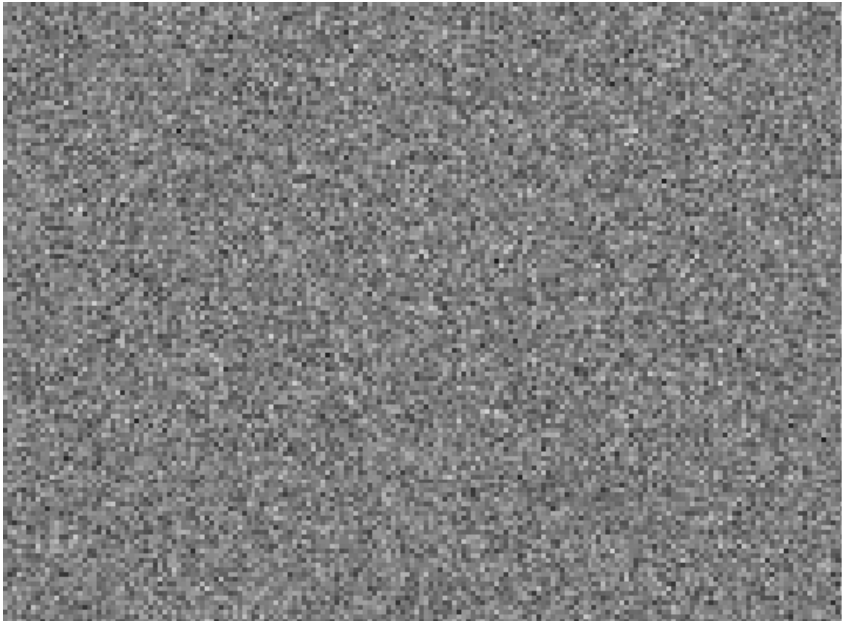}
    \caption{Simulation without correlated noise between channels.}
    \label{fig:withoutM}
  \end{subfigure}
  \begin{subfigure}{0.45\textwidth}
    \centering
    \includegraphics[width=\linewidth]{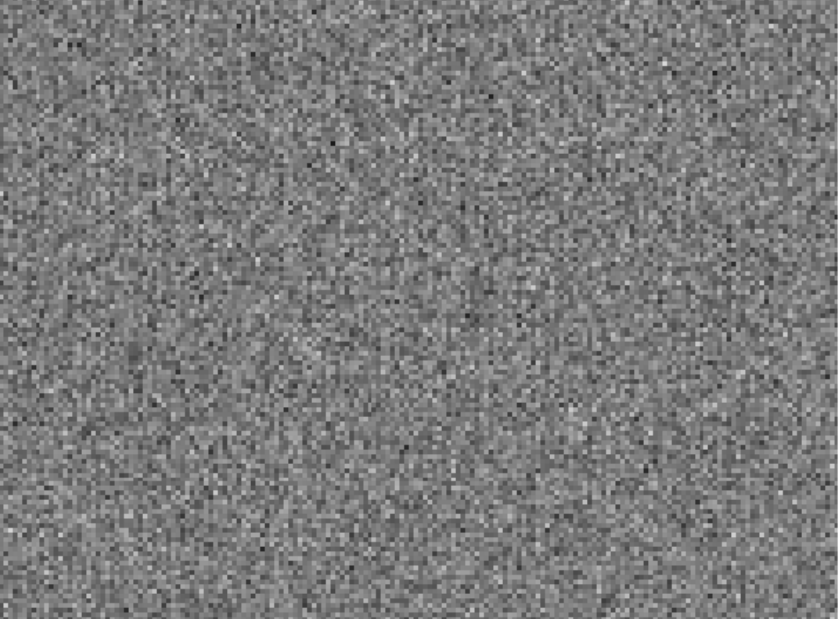}
    \caption{Simulation adds joint wide-sense stationary.}
    \label{fig:withM}
  \end{subfigure}

  \caption{A portion of the overscan from the first channel ($i=1$) of the simulated images with $\text{NSAMP}=1$, a readout noise of $\sigma = 4\,e^-$ for all channels, a gain of $G = 50\,\text{ADU}/e^-$, and $\mathrm{SC} = 10^{-4}\,e^-/\text{pix}/\text{transfer}$.}
  \label{fig:image_sample}
\end{figure}

Figure \ref{fig:image_sample} shows a portion of the overscan region of images for the first amplifier. Figure \ref{fig:withoutM} was simulated using a readout noise of $4\,e^-$ per amplifier, a horizontal SC of $10^{-4}\,e^-/\text{pix}/\text{transfer}$, and no correlated noise. Figure \ref{fig:withM} adds an extra jointly wide-sense stationary noise component affecting all channels. This extra noise was modeled as
\begin{equation*}
m_i(n) = A \sin\!\left(\frac{3n}{2\pi} + B\right), \quad \forall\, i,\; 1 \le i \le 16,
\end{equation*}
where $A = 1\,e^-$ and $B$ is a random variable. In this case, the coupling of each channel to the noise is the same. The noise amplitude was set to be small relative to the readout noise of each channel. This correlated noise produced the subtle vertical lines observed in Figure \ref{fig:withM}.

\begin{figure}[h!]
  \centering
  \begin{subfigure}{0.40\columnwidth}
    \centering
    \includegraphics[width=1\textwidth]{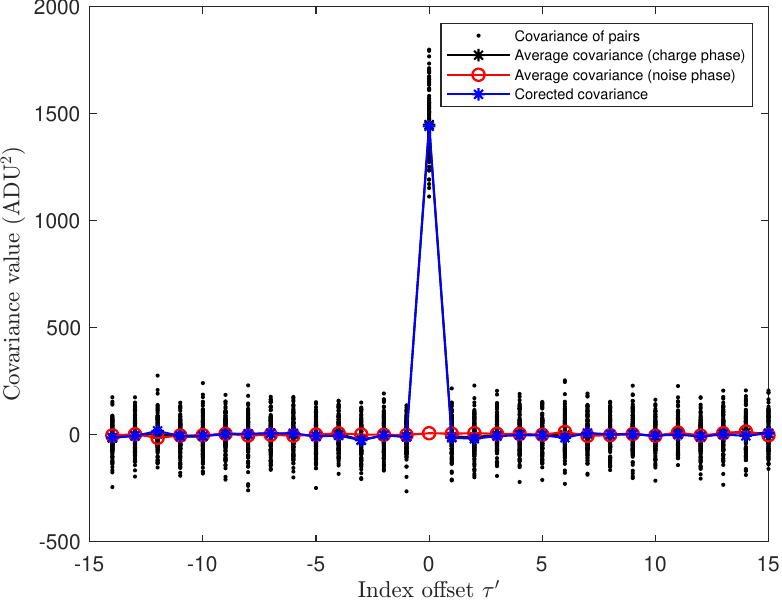}
    \caption{$\mathrm{SC} = 10^{-3} e^-/\text{pix}/\text{transfer}$ and no correlated noise between channels.}
    \label{fig:covariance-10e-3}
  \end{subfigure}
  \begin{subfigure}{0.40\columnwidth}
    \centering
    \includegraphics[width=1\textwidth]{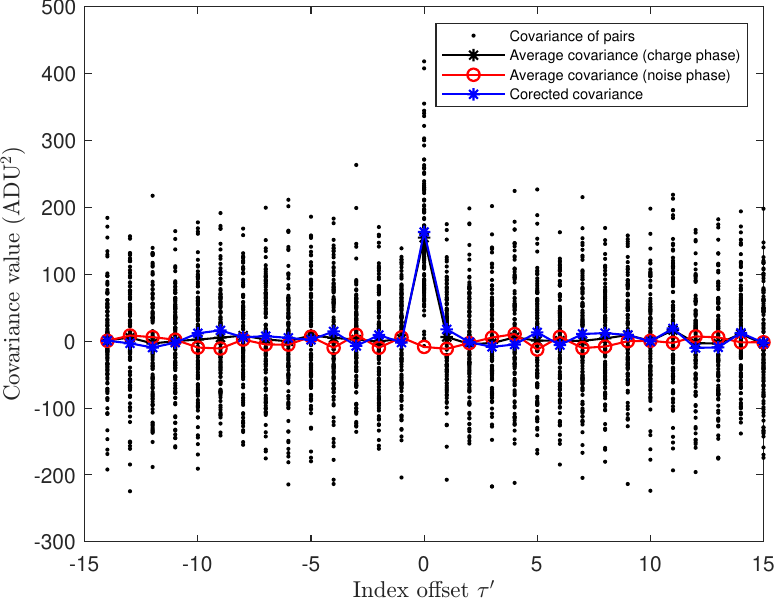}
    \caption{$\mathrm{SC} = 10^{-4} e^-/\text{pix}/\text{transfer}$ and no correlated noise between channels.}
    \label{fig:covariance-10e-4}
  \end{subfigure}
  \begin{subfigure}{0.40\columnwidth}
    \centering
    \includegraphics[width=1\textwidth]{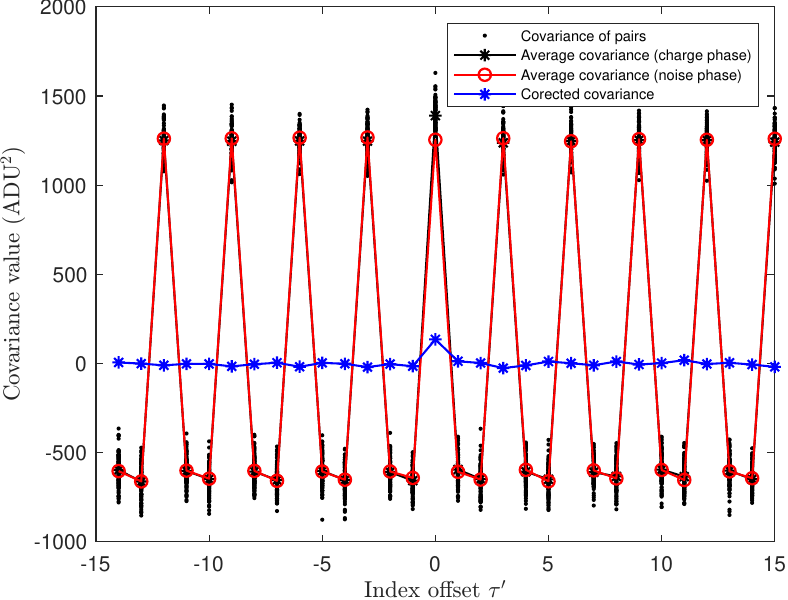}
    \caption{Simulation with joint wide-sense stationary noise between channels.}
    \label{fig:covariance-10e-4-withM}
  \end{subfigure}

  \caption{The images are simulated using $\text{NSAMP} = 1$, a readout noise of $\sigma=4 \, e^-$ for all channels, and a gain $G=50 \text{ ADU}/e^-$.}
  \label{fig:covariances-plot}
\end{figure}

Figure \ref{fig:covariances-plot} shows the calculated covariance amplitude, following the calculations in Eq. (\ref{eq: cov estimator}), as a function of the pixel difference between channels. The index offset, $\tau'$, on the abscissa axis is the additional offset with respect to the charge phase offset, i.e., $\tau=(j-i)N_p + \tau'$. This definition makes it possible to combine the correlations from all channels using the same indexing. The black dots show the covariance for each pair of channels, $c_{i,j}$, and the black stars denote their simple average. The red circles represent the average correlation across all channels in the noise phase, using the index offset $\tau^M = (i-j)N_p + \tau'$. Finally, the blue dots show the corrected covariance in the charge phase after subtracting the contribution from correlated noise, as shown in Eq. (\ref{eq:cov correctoin}).

Figure~\ref{fig:covariance-10e-3} shows a case with relatively high SC of $10^{-3}\,e^-/\text{pix}/\text{transfer}$. There is no correlated noise in this case, so the red dots are around zero for all indices. When the index is different from zero, the channels are not in charge phase, and the correlation among channels is also around zero. In Figure~\ref{fig:covariance-10e-4}, the SC level is $10^{-4}\,e^-/\text{pix}/\text{transfer}$, so the values at $\tau'=0$ are smaller than in the previous case by approximately one order of magnitude. Despite this, for small SC values, the technique still allows the detection of the extra correlation above the floor imposed by the electronic noise. The third case, shown in Figure~\ref{fig:covariance-10e-4-withM}, is the same as in Fig.~\ref{fig:covariance-10e-4} but with the addition of correlated noise. In this case, the correlated noise contribution to the covariance is high; in particular, its contribution is greater than that of the charge signal. The black dots and the black stars from the covariance in the charge phase follow the covariance in the noise phase for all indices. This highlights the symmetry exhibited by stationary noise. At $\tau'=0$, after correcting the covariance, it is possible to recover the expected value of the charge contribution. This demonstrates the relevance of a technique that allows processing the data even in realistic scenarios where the correlated noise cannot be eliminated beforehand by preprocessing the images.

\begin{figure}[h!]
    \centering
    \includegraphics[width=0.55\linewidth]{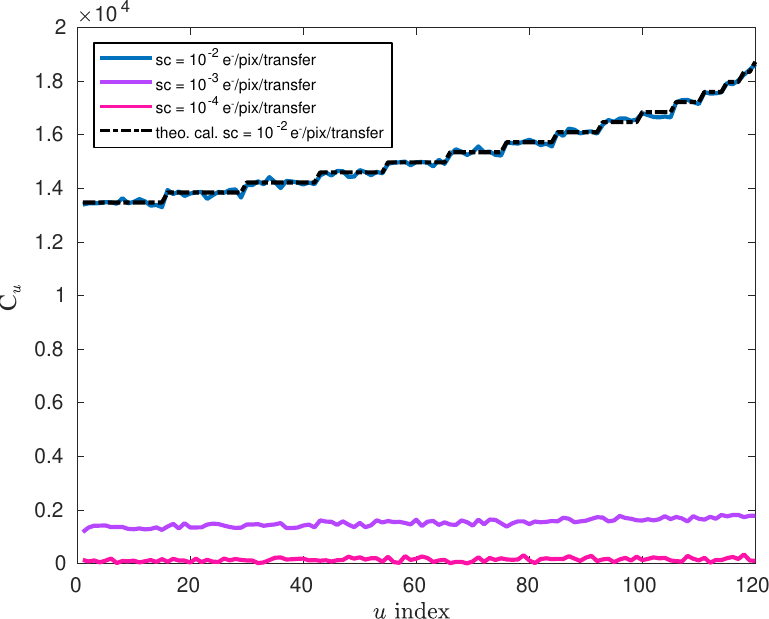}
    \caption{$C_u$ estimated for different conditions of SC. The images are simulated using $\text{NSAMP} = 1$, readout noise of $\sigma=4 \, e^-$ for all channels, and gain $G=50\text{ ADU}/e^-$.}
    \label{fig:cu}
\end{figure}

Figure \ref{fig:cu} summarizes the covariance in Eq. (\ref{eq:cov-estimator-cp}) for all channels evaluated at $\tau'=0$. For convenience, we use the $u$ index defined in Eq. (\ref{u index}) to uniquely define each value. The colored curves are calculations from the simulated images for different levels of SC. The cyan curve shows the estimated $C_u$ values for $\mathrm{SC}=10^{-2} e^-/\text{pix}/\text{transfer}$. The piecewise-constant behavior observed here and in the theoretical dashed curve is derived from Eq. (\ref{eq: covij}), as the covariance of a pair always reflects the SC collected until the lower-index amplifier. Thus, these steps are progressively smaller and higher in amplitude toward the right of the curve as a result of the extra transfers required to reach those amplifiers. The purple curve shows the covariance when $\mathrm{SC}=10^{-3} e^-/\text{pix}/\text{transfer}$. The curve shows a small increase in the covariance at higher indices, but the staggered pattern is more difficult to visualize due to a smaller SC value. Finally, the pink curve shows the covariance when $\mathrm{SC}=10^{-4} e^-/\text{pix}/\text{transfer}$. The covariance value is much smaller. An interesting aspect to note is that, for the three cases, the small random fluctuations in the calculations have a similar amplitude. As shown in the variance calculation in Eq. (\ref{eq:estimator-variance}), this is due to the fact that, for small SC values, the error in covariance estimation is primarily determined by the readout noise in the channels.

\subsection{Covariance weighting for spurious charge calculation}

\begin{figure}[h!]
  \centering
  \begin{subfigure}{0.40\columnwidth}
    \centering
    \includegraphics[width=1\linewidth]{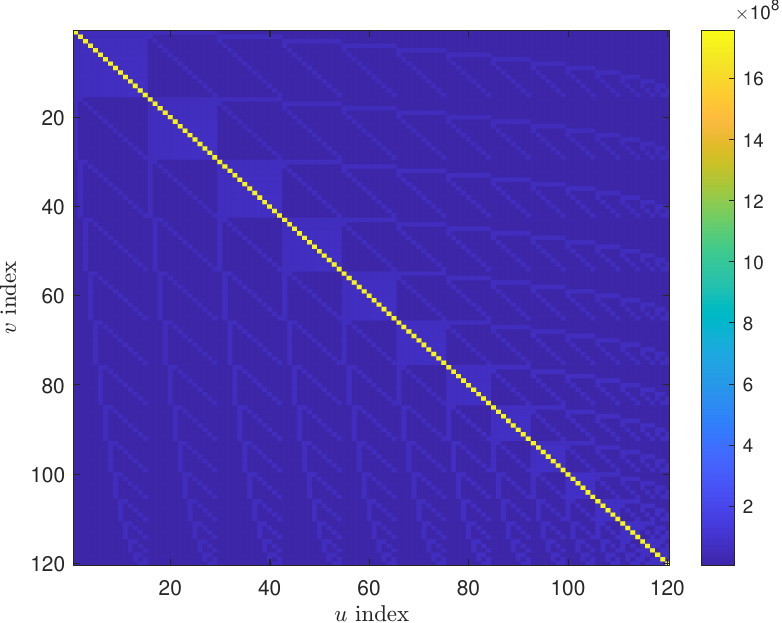}
    \caption{$\mathrm{SC} = 10^{-3} e^-/\text{pix}/\text{transfer}$ and no correlated noise between channels.}
    \label{fig:sigma-matrix-a}
  \end{subfigure}
  \begin{subfigure}{0.40\columnwidth}
    \centering
    \includegraphics[width=1\linewidth]{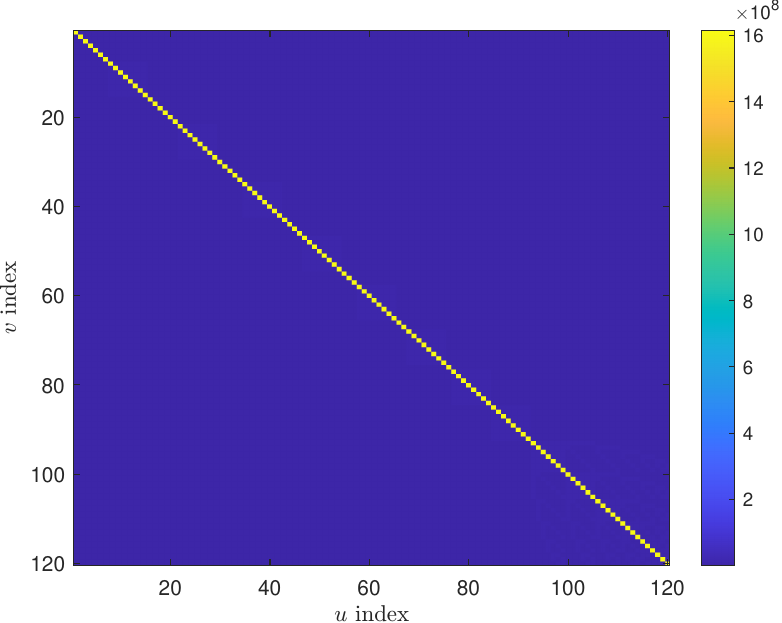}
    \caption{$\mathrm{SC} = 10^{-4} e^-/\text{pix}/\text{transfer}$ and no correlated noise between channels.}
    \label{fig:sigma-matrix-b}
  \end{subfigure}
  \begin{subfigure}{0.40\columnwidth}
    \centering
    \includegraphics[width=1\linewidth]{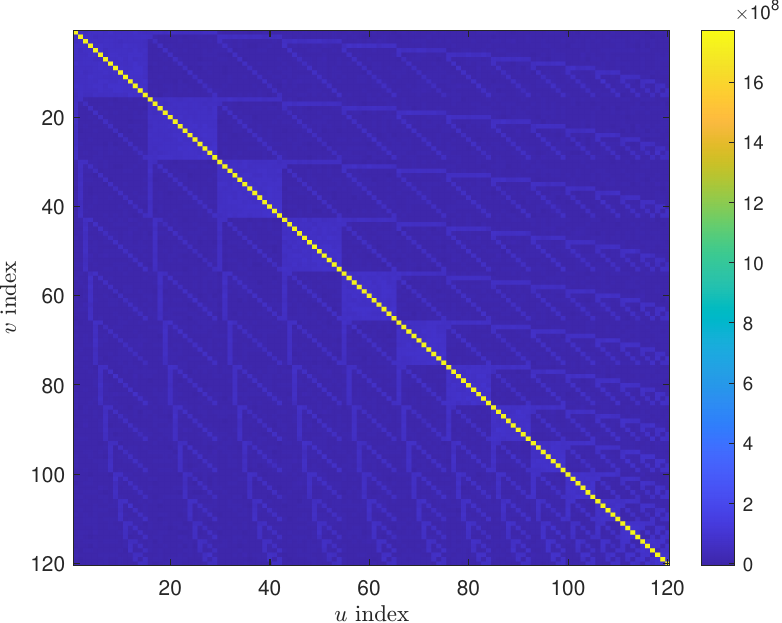}
    \caption{$\mathrm{SC} = 10^{-4} e^-/\text{pix}/\text{transfer}$ with joint wide sense stationary noise between channels.}
    \label{fig:sigma-matrix-c}
  \end{subfigure}

  \caption{$\Sigma$ matrices calculated for different conditions of SC and correlated noise between channels. The matrices are calculated using $\text{NSAMP}=1$, a readout noise of $\sigma=4e^-$ for all channels, and a gain $G=50 \text{ADU}/e^-$.}
  \label{fig:sigma-matrix}
\end{figure}

The next step in the calculation process involves estimating the SC values from the simulated images based on the covariance analysis. Figure \ref{fig:sigma-matrix} shows, on a color scale, the values of the $\Sigma$ matrix from Eq. (\ref{eq:Sigma matrix}) obtained for three different conditions of SC and noise. In Figure \ref{fig:sigma-matrix-a}, a SC of $10^{-3} e^-/\text{pix}/\text{transfer}$ and a readout noise per channel of $4e^-$ was simulated. The largest values are the elements in yellow on the main diagonal. The main contribution of these elements is the first term of Eq. (\ref{eq:cov-expression}) given by the product of the readout noise variance of each channel used in the covariance calculation. There are other non-diagonal elements with values greater than the minimum. These are primarily determined by other terms in the equation, which involve the multiplication of charge values and channel readout noise. When the SC is small, as in the case of Figure \ref{fig:sigma-matrix-b}, with $\mathrm{SC}=10^{-4} e^-/\text{pix}/\text{transfer}$, the non-diagonal elements have less relative amplitude. In this case, when the SC is very small, $\Sigma$ suggests that there is relatively small cross-information between the covariance estimators, and therefore the estimation using the simple weights from Eq. (\ref{eq:simple-weights}) should give a result close to the optimal solution in Eq. (\ref{eq:min_var}). When correlated noise is incorporated, as in the case of Figure \ref{fig:sigma-matrix-c}, the non-diagonal elements regain relevance. In this case, the optimal solution might be worth computing if no decorrelation is possible.

\begin{figure}[h!]
  \centering
  \begin{subfigure}{0.45\columnwidth}
    \centering
    \includegraphics[width=1\linewidth]{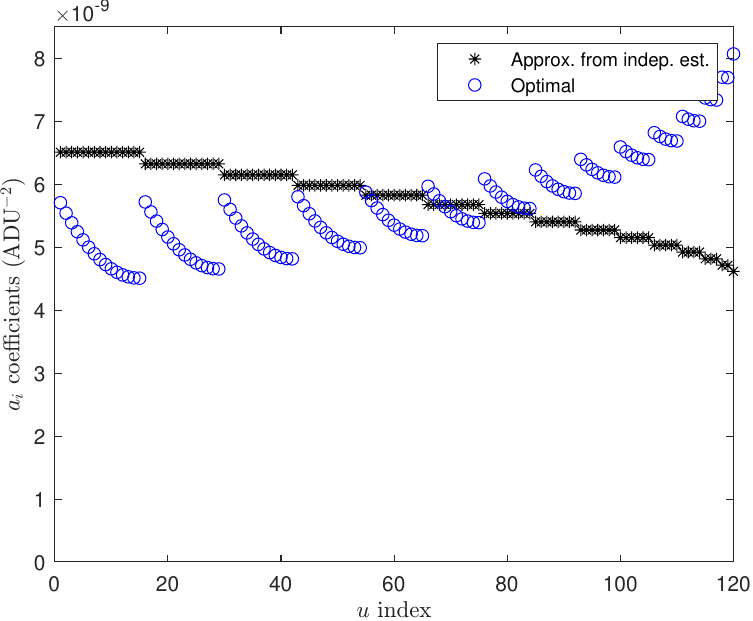}
    \caption{$\mathrm{SC} = 10^{-3} e^-/\text{pix}/\text{transfer}$.}
    \label{fig:weights-a}
  \end{subfigure}
  \begin{subfigure}{0.45\columnwidth}
    \centering
    \includegraphics[width=1\linewidth]{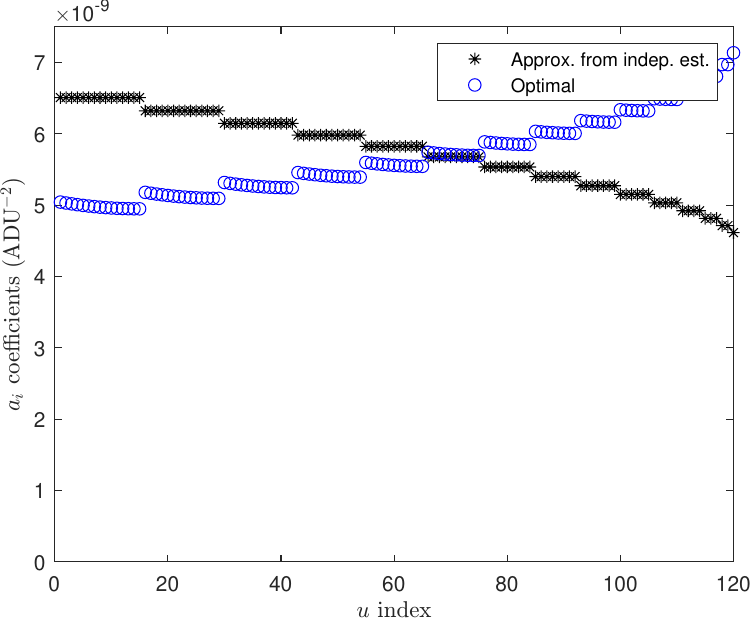}
    \caption{$\mathrm{SC} = 10^{-4} e^-/\text{pix}/\text{transfer}$.}
    \label{fig:weights-b}
  \end{subfigure}

  \caption{Calculated coefficients for the matrices calculated for different conditions of SC and correlated noise between channels. The matrices are calculated using $\text{NSAMP}=1$, a readout noise of $\sigma=4e^-$ for all channels, gain $G=50 \text{ADU}/e^-$ and no correlated noise between channels.}
  \label{fig:weights}
\end{figure}

Figure \ref{fig:weights} shows the calculated weights used to combine the covariances under the two different SC conditions. The blue circles correspond to the optimal weights from Eq. (\ref{eq:optimal-solution}), while the black dots denote the simple weights from Eq.~(\ref{eq:simple-weights}). Since this formula depends only on the distance vector $\mathbf{D}$—which quantifies the extra number of transfers to reach the successive amplifiers—the SC value and correlated noise have no effect. Consequently, both plots show identical weights. Values are smaller for greater amplifier indices to counterbalance the extra number of transfers and provide an average estimate value of SC. The flat interval values correspond to weights associated with covariances calculated using a common amplifier. Since the correlation in charge phase between two amplifiers is given by the charge produce before the first amplifier in the chain, those covariances provide similar charge information and are weighted equally.

In the case of the optimal calculation of weights, their value is calculated with extra information on the statistical errors of the covariance estimation, so the general shape of the weight series is slightly different. They show a general growing trend for larger values towards higher indices, given by the fact that the last amplifiers in the chain have larger values of SC per pixel (due to the extra transfers) and therefore a larger signal-to-noise ratio to calculate the SC value. Within each of the continuous segments, the value of the weights is not flat. This is explained because the estimation of the covariance has larger errors for amplifiers that are further apart, since the extra charge generated between them adds uncertainty to the estimation. This behavior is observed in higher values of the left weights on each of the blue segments. This is more evident for the blue dots in Figure \ref{fig:weights-a}, which were simulated with a larger contribution of SC. For a smaller contribution of SC, as in Figure \ref{fig:weights-b}, this is less evident. 

\subsection{Estimation of the spurious charge value}

\begin{figure}[h!]
    \centering
    \includegraphics[width=0.45\linewidth]{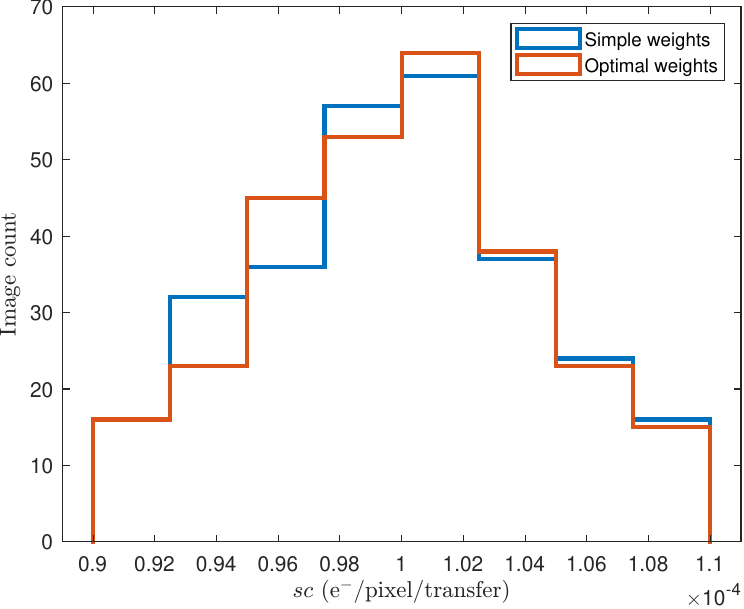}
    \caption{Histogram of the estimated values of SC using the proposed technique. The blue histogram is calculated using the approximate solution for the weights to average the covariance values from Eq. (\ref{eq:simple-weights}). The orange histogram is using the optimal weights from Eq. (\ref{eq:optimal-solution}).}
    \label{fig:histogram of calculated values}
\end{figure}

Figure \ref{fig:histogram of calculated values} shows two histograms for the estimated SC values obtained from 293 simulated images. The simulation conditions are: $\sigma = 4e^-$ readout noise per channel, $N_a = 16$ output amplifiers separated by $N_p = 15$ pixels, $N_{\text{act}} = 512$ active columns, no correlated noise, and a SC level of $10^{-4}\, e^-/\text{pix}/\text{transfer}$. The light-blue histogram is obtained using simple weights that arise from the distance vector, $\mathbf{D}$, from Eq.~(\ref{eq:decomposition}), while the orange histogram corresponds to the optimal solution from Eq.~(\ref{eq:optimal-solution}). For this case, both approaches yield similar results, with both histograms centered around the simulated SC level and showing small dispersion. When the SC is small, the covariance matrix, $\Sigma$, is dominated by its diagonal terms. Therefore, the optimal weights do not differ significantly from the simplified solution.

The sensitivity of the technique depends on several factors: the chosen region of interest, the number of transfers each pixel undergoes, the number of output amplifiers, and the readout noise of each amplifier. Using the variance of the estimator from Eq.~(\ref{eq:estimator-variance}) and under the assumption that all channels have similar noise, $\sigma$, the number of columns in the active region of the sensor is much larger than the extended serial register, $N_{\text{act}} \gg (N_a-1)N_p$, and the SC to be estimated is small such that $\lambda_{i} = \text{SC}(N_{\text{act}}+(i-1)N_p) \ll \sigma$, the error in the estimation of the SC value, using a portion of the image with $N_{\text{pix}}$ pixels, expressed in units of
$e^-/\text{pix}/\text{transfer}$, is 
\begin{equation}
 \sigma_{\mathrm{SC}} = \frac{\sigma^2}{N_{\text{act}}\sqrt{N_c N_{\text{pix}}}}.
 \label{eq:method-precission}
\end{equation}
Replacing a common case used across sections, with $\sigma = 4e^-$, $N_{\text{act}} = 512$, $N_a = 16$, and $N_{\text{pix}} = 3\times10^5$, the previous equation yields $\sigma_{\mathrm{SC}} = 5.2\times10^{-6}\, e^-/\text{pix}/\text{transfer}$.

From Eq. (\ref{eq:method-precission}), the possibility to define a desired precision by choosing a specific region of interest derives:
\begin{equation}
  N_{\text{pix}}= \frac{\sigma^4}{N_c (N_{\text{act}}\sigma_{\text{SC}})^2}.
 \label{eq:precission of the method}
\end{equation}
For example, to achieve a precision of $\sigma_{\text{SC}}=10^{-5}\, e^-/\text{pix}/\text{transfer}$ with a sensor having $N_{\text{act}} = 2000$, a readout noise of $\sigma=3e^-$, and $N_a = 16$ output amplifiers, the required number of pixels is $N_{\text{pix}} = 1688$.

\subsection{Comparison with other methods}

Finally, the proposed technique was compared to other existing methods to measure SC in the horizontal overscan region. One of the widely used techniques is based on measuring the change in the probability distribution \cite{Villalpando_2024}. Typically, this method involves fitting a probability density function to the data. This function is usually the convolution of the normal density function, which represents the statistics of the electronic noise, with a Poisson density function, which models the charge in the pixels produced by the SC. To simplify the calculations, the simulations presented below consider only two output amplifiers. For the fitting procedure, the information from both amplifiers is averaged to obtain a final image with lower readout noise. For the SC technique presented here, the individual information from each amplifier is used to estimate a single covariance value. $100$ images were simulated for each SC level, with a region size of $6000 \times 50$ pixels, readout noise of $\sigma = 4e^-$ for a sensor with the same characteristics as those mentioned in Section \ref{sec:simulations}.

\begin{figure}[h!]
  \centering
  \begin{subfigure}{0.45\columnwidth}
    \centering
    \includegraphics[width=0.99\linewidth]{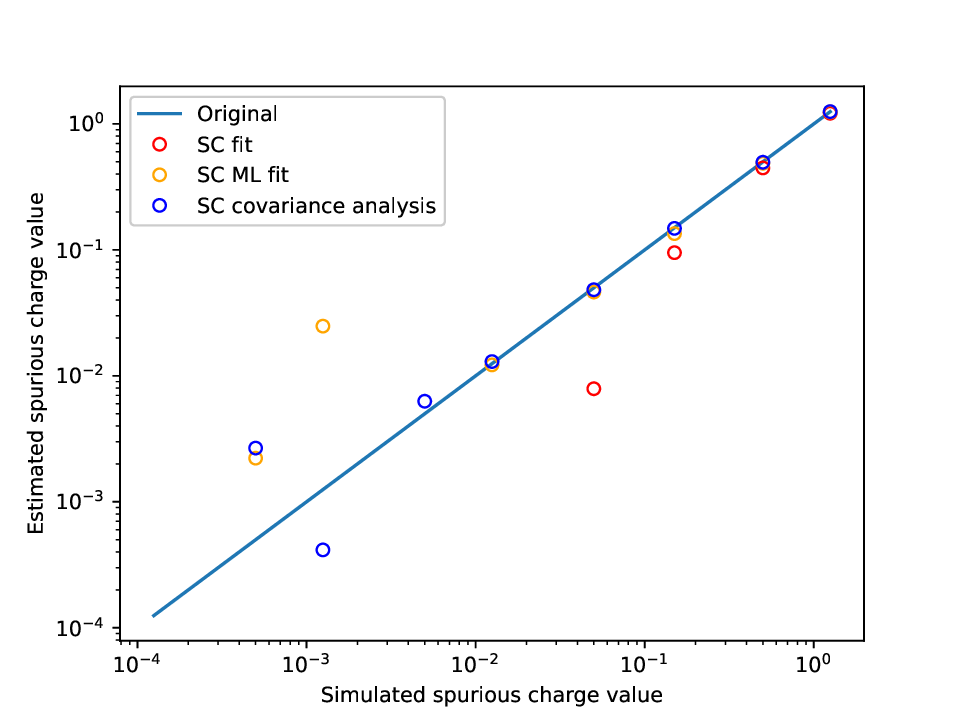}
    \caption{Simulated and estimated value comparison.}
    \label{fig:compa}
  \end{subfigure}
  \begin{subfigure}{0.45\columnwidth}
    \centering
    \includegraphics[width=0.99\linewidth]{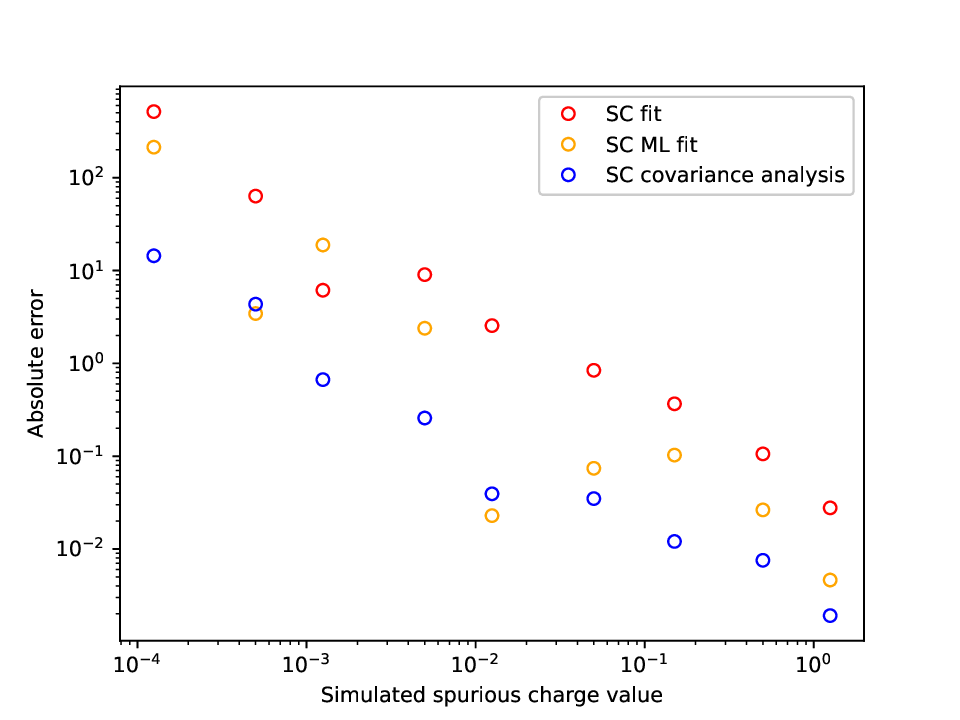}
    \caption{Error of each method for each SC simulated.}
    \label{fig:compb}
  \end{subfigure}

      \caption{Performance comparison of the proposed technique with other methods typically used to estimate SC.}
  \label{fig:methods comparison}
\end{figure}

Figure \ref{fig:compa} shows the simulated values on the abscissa axis and the estimated SC values on the ordinate axis. For the estimations, three approaches are applied: a conventional binned $\chi^2$ fit (red circles); a finer binned fit with ten times more bins combined with a maximum likelihood method (orange circles), which is expected to improve upon the previous approach; and the covariance method (blue) presented in this work. At high SC values, all three methods give precise estimates of the simulated SC, but when the SC is reduced, the error increases. The basic binned fit gives the largest discrepancies, while the covariance approach provides a better estimate. In contrast, Figure \ref{fig:compb} shows the relative error as a function of the simulated SC, confirming that the worst-performing method is the basic binned fit, the improved binned fit achieves better agreement with the simulated data, and the best results are obtained using the covariance approach. This establishes covariance analysis as a practical instrument for SC characterization in MAS-CCD devices.

\section{Conclusion}

This work highlights the importance of studying sub-electron levels of SC as a non-negligible noise source for single-electron-resolution imaging sensors. Currently, available methods are not capable of precisely measuring such small contributions of SC at relatively short readout times, motivating the development of a different measurement technique to study this noise source in this regime. This work introduces a covariance-based framework that leverages the MAS-CCD architecture to estimate SC directly from standard readout images. This approach enables fast and reliable SC estimation without necessarily relying on single-electron-resolution measurements. The method provides accurate estimation under realistic noise regimes and performs competitively with other available techniques. Its performance remains robust even in the presence of correlated, jointly wide-sense stationary noise. With the growing interest in MAS-CCD devices for astronomical applications, this technique opens the possibility of fast and systematic SC characterization over a wide range of clock swings and other operating conditions, such as timing configurations or samples per pixel.

\section*{Acknowledgment}
The multi-amplifier sensing (MAS) CCD was developed as a collaborative endeavor between Lawrence Berkeley National Laboratory and Fermi National Accelerator Laboratory. Funding for the design and fabrication of the MAS device described in this work came from a combination of sources including the DOE Quantum Information Science (QIS) initiative, the DOE Early Career Research Program, and the Laboratory Directed Research and Development Program at Fermi National Accelerator Laboratory under Contract No. DE-AC02-07CH11359. This research has been partially supported by the Heising-Simons Foundation (Grant No. 2023-4611) and Guillermo Fernandez Moroni’s DOE Early Career research program.

This manuscript has been authored by Fermi Forward Discovery Group, LLC under Contract No. 89243024CSC000002 with the U.S. Department of Energy, Office of Science, Office of High Energy Physics. 

\appendix
\section{Mathematical Derivations} 
\label{ap:derivations}

In this appendix, a general form for the variance of the covariance estimator is presented. The covariance between pixel products $C_u$ and $C_v$ can be computed as:

\begin{equation}
\label{eq:cov-1}
\begin{aligned}
\mathrm{Cov}(C_u,C_v)
&= \mathrm{E}\!\left[
    (\qz{P}{i}\qz{P}{j}-\mathrm{E}[\qz{P}{i}\qz{P}{j}])
    (\qz{P}{k}\qz{P}{l}-\mathrm{E}[\qz{P}{k}\qz{P}{l}])
  \right] \\
&= \mathrm{E}\!\left[
    ((\qz{R}{i}+\qz{Q}{i})(\qz{R}{j}+\qz{Q}{j})-\lambda_i)
    ((\qz{R}{k}+\qz{Q}{k})(\qz{R}{l}+\qz{Q}{l})-\lambda_k)
  \right] \\
&= \mathrm{E}\!\left[
    (\qz{R}{i}(\qz{R}{j}+\qz{Q}{j})
     +\qz{Q}{i}(\qz{R}{j}+\qz{Q}{j})
     -\lambda_i)
\right.\\
&\qquad\left.
    (\qz{R}{k}(\qz{R}{l}+\qz{Q}{l})
     +\qz{Q}{k}(\qz{R}{l}+\qz{Q}{l})
     -\lambda_k)
  \right] \\
&= \mathrm{E}\!\left[
    (\qz{R}{i}\qz{R}{j}
     +\qz{R}{i}\qz{Q}{j}
     +\qz{Q}{i}\qz{R}{j}
     +\qz{Q}{i}\qz{Q}{j}
     -\lambda_i)
\right.\\
&\qquad\left.
    (\qz{R}{k}\qz{R}{l}
     +\qz{R}{k}\qz{Q}{l}
     +\qz{Q}{k}\qz{R}{l}
     +\qz{Q}{k}\qz{Q}{l}
     -\lambda_k)
  \right].
\end{aligned}
\end{equation}
where the indices $u = f(i,j), \ i<j $ and $v = f(k,l), \ k<l$ were defined according to the unambiguous index redefinition given in Eq. (\ref{u index}). Distributing the terms inside the expectation produces 25 terms, of which 16 evaluate to zero due to the independence among the random variables. Thus, Eq. (\ref{eq:cov-1}) reduces to:
\begin{equation}
\label{eq:full-covariance-2}
\begin{aligned}
\mathrm{Cov}(C_u,C_v)
&= \mathrm{E}\Big[
    \qz{R}{i}\qz{R}{j}\qz{R}{k}\qz{R}{l} \\
&\quad
  + \qz{R}{i}\qz{Q}{j}\qz{R}{k}\qz{Q}{l}
  + \qz{R}{i}\qz{Q}{j}\qz{Q}{k}\qz{R}{l} \\
&\quad
  + \qz{Q}{i}\qz{R}{j}\qz{R}{k}\qz{Q}{l}
  + \qz{Q}{i}\qz{R}{j}\qz{Q}{k}\qz{R}{l} \\
&\quad
  + \qz{Q}{i}\qz{Q}{j}\qz{Q}{k}\qz{Q}{l} \\
&\quad
  - \qz{Q}{i}\qz{Q}{j}\lambda_k
  - \lambda_i\qz{Q}{k}\qz{Q}{l}
  + \lambda_i\lambda_k
\Big].
\end{aligned}
\end{equation}

So far, the conditions on indices $i$, $j$, $k$, and $l$ have been stated but no restriction on their order has been specified. However, this turns out to be relevant because the order determines what the expected value evaluates to. For example, in the second term of Eq. (\ref{eq:full-covariance-2}), there are three possibilities:
\begin{equation*}
    \mathrm{E} \big[\qz{R}{i} \qz{R}{k} \qz{Q}{j} \qz{Q}{l} \big] = 
    \begin{cases}
        \sigma_{i}^2 \lambda_{j}, & \text{if } i = k \ \text{and} \ j < l, \\
        \sigma_{i}^2 \lambda_{l}, & \text{if } i = k \ \text{and} \ l < j, \\
        0, & \text{elsewhere}.
    \end{cases}
\end{equation*}
In general, terms having a pair of $\tilde{Q}$ variables evaluate to the $\lambda$ corresponding to the smallest index, whereas terms that have a pair of $\tilde{R}$ variables are null unless the indices match. To obtain a more general formulation that allows one to derive easier-to-handle expressions, an index redefinition is introduced: $h_1$, $h_2$, $h_3$, and $h_4$ are defined such that $h_1 \leq h_2 \leq h_3 \leq h_4$. This quartet takes the values of $i$, $j$, $k$, and $l$ according to their ascending order, e.g., if $(i,j,k,l) = (1, 4, 2, 3)$ then $(h_1, h_2, h_3, h_4) = (i, k, l, j)$. Under this redefinition, the former expression can be simplified to
\begin{equation*}
    \mathrm{E} \big[\qz{R}{h_1} \qz{R}{h_3} \qz{Q}{h_2} \qz{Q}{h_4} \big] = \delta(h_1 - h_3) \sigma_1^2 \lambda_{h_2},
\end{equation*}
where $\delta(x)$ is the Dirac delta function, which is nonzero only when $x = 0$. Reformulating Eq. (\ref{eq:full-covariance-2}) results in

\begin{equation*}
\begin{aligned}
\mathrm{Cov}(C_u,C_v)
&= \mathrm{E}\Big[
    \qz{R}{h_1}\qz{R}{h_2}\qz{R}{h_3}\qz{R}{h_4} \\
&\quad
  + \qz{R}{h_1}\qz{Q}{h_2}\qz{R}{h_3}\qz{Q}{h_4}
  + \qz{R}{h_1}\qz{Q}{h_2}\qz{Q}{h_3}\qz{R}{h_4} \\
&\quad
  + \qz{Q}{h_1}\qz{R}{h_2}\qz{R}{h_3}\qz{Q}{h_4}
  + \qz{Q}{h_1}\qz{R}{h_2}\qz{Q}{h_3}\qz{R}{h_4} \\
&\quad
  + \qz{Q}{h_1}\qz{Q}{h_2}\qz{Q}{h_3}\qz{Q}{h_4} \\
&\quad
  - \qz{Q}{h_1}\qz{Q}{h_2}\lambda_{h_3}
  - \lambda_{h_1}\qz{Q}{h_3}\qz{Q}{h_4}
  + \lambda_{h_1}\lambda_{h_3}
\Big] \\
&= \mathrm{E}[\qz{R}{h_1}\qz{R}{h_2}\qz{R}{h_3}\qz{R}{h_4}] \\
&\quad
  + \mathrm{E}[\qz{R}{h_1}\qz{R}{h_3}]\,\lambda_{h_2}
  + \mathrm{E}[\qz{R}{h_1}\qz{R}{h_4}]\,\lambda_{h_2} \\
&\quad
  + \mathrm{E}[\qz{R}{h_2}\qz{R}{h_3}]\,\lambda_{h_1}
  + \mathrm{E}[\qz{R}{h_2}\qz{R}{h_4}]\,\lambda_{h_1} \\
&\quad
  + \mathrm{E}[\qz{Q}{h_1}\qz{Q}{h_2}\qz{Q}{h_3}\qz{Q}{h_4}]
  - \lambda_{h_1}\lambda_{h_3}.
\end{aligned}
\end{equation*}
where each pair of charge variables was replaced with its corresponding $\lambda$ parameters. Using the previously mentioned property of the $\tilde{R}$ variables, the final form of the covariance reduces to:
\begin{equation}\label{eq:full-cov-expression}
\begin{aligned}
\text{Cov}(C_u, C_v)
&= \delta(h_1 - h_3) \delta(h_2 - h_4) \sigma_{h_1}^2 \sigma_{h_2}^2 \\
&\quad + \delta(h_1 - h_3) \sigma_{h_1}^2 \lambda_{h_2}
      + \delta(h_1 - h_4) \sigma_{h_1}^2 \lambda_{h_2} \\
&\quad + \delta(h_2 - h_3) \sigma_{h_1}^2 \lambda_{h_1}
      + \delta(h_2 - h_4) \sigma_{h_2}^2 \lambda_{h_1} \\
&\quad - \lambda_{h_1} \lambda_{h_2}
      + \mathrm{E}[\qz{Q}{h_1}\qz{Q}{h_2}\qz{Q}{h_3}\qz{Q}{h_4}].
\end{aligned}
\end{equation}
Note that, as discussed in Section \ref{sec: optimal weights}, Eq. (\ref{eq:full-cov-expression}) matches the form presented in Eq. (\ref{eq:cov-expression}). The last term of the covariance requires further analysis. It can be rewritten using the variable defined in Eq. (\ref{eq:extra-variableQij}), which represents the charge generated between two amplifiers, as:
\begin{equation*}
\begin{aligned}
\mathrm{E}[\qz{Q}{h_1}\qz{Q}{h_2}\qz{Q}{h_3}\qz{Q}{h_4}]
&= \mathrm{E}\big[
    \qz{Q}{h_1}(\qz{Q}{h_1}+\qz{Q}{h_1,h_2}) \\
&\quad
    (\qz{Q}{h_1}+\qz{Q}{h_1,h_2}+\qz{Q}{h_2,h_3}) \\
&\quad
    (\qz{Q}{h_1}+\qz{Q}{h_1,h_2}+\qz{Q}{h_2,h_3}+\qz{Q}{h_3,h_4})
\big].
\end{aligned}
\end{equation*}
Then, after distributing all terms, the full expanded expected value is:
\begin{multline*}
\mathrm{E}\big[
\qz{Q}{h_1}^4
+ \qz{Q}{h_1}^3 \qz{Q}{h_1,h_2}
+ \qz{Q}{h_1}^3 \qz{Q}{h_2,h_3}
+ \qz{Q}{h_1}^3 \qz{Q}{h_3,h_4}
+ 2 \qz{Q}{h_1}^3 \qz{Q}{h_1,h_2}
+ 2 \qz{Q}{h_1}^2 \qz{Q}{h_1,h_2}^2 \\
+ 2 \qz{Q}{h_1}^2 \qz{Q}{h_1,h_2} \qz{Q}{h_2,h_3}
+ 2 \qz{Q}{h_1}^2 \qz{Q}{h_1,h_2} \qz{Q}{h_3,h_4}
+ \qz{Q}{h_1}^3 \qz{Q}{h_2,h_3}
+ \qz{Q}{h_1}^2 \qz{Q}{h_1,h_2} \qz{Q}{h_2,h_3}
+ \qz{Q}{h_1}^2 \qz{Q}{h_2,h_3}^2 \\
+ \qz{Q}{h_1}^2 \qz{Q}{h_2,h_3} \qz{Q}{h_3,h_4}
+ \qz{Q}{h_1}^2 \qz{Q}{h_1,h_2}^2
+ \qz{Q}{h_1} \qz{Q}{h_1,h_2}^3
+ \qz{Q}{h_1} \qz{Q}{h_1,h_2}^2 \qz{Q}{h_2,h_3}
+ \qz{Q}{h_1} \qz{Q}{h_1,h_2}^2 \qz{Q}{h_3,h_4} \\
+ \qz{Q}{h_1}^2 \qz{Q}{h_1,h_2} \qz{Q}{h_2,h_3}
+ \qz{Q}{h_1} \qz{Q}{h_1,h_2}^2 \qz{Q}{h_2,h_3}
+ \qz{Q}{h_1} \qz{Q}{h_1,h_2} \qz{Q}{h_2,h_3}^2
+ \qz{Q}{h_1} \qz{Q}{h_1,h_2} \qz{Q}{h_2,h_3} \qz{Q}{h_3,h_4}
\big].
\end{multline*}
Finally, the independence property given in Eq. (\ref{eq: indep of charge}) helps to reduce the expression to its final form:
\begin{equation}
\begin{aligned}
\mathrm{E}[\qz{Q}{h_1} \qz{Q}{h_2} \qz{Q}{h_3} \qz{Q}{h_4}]
&= \mathrm{E}[\qz{Q}{h_1}^4]
 + 3\,\mathrm{E}[\qz{Q}{h_1}^2 \qz{Q}{h_1,h_2}^2]
 + \mathrm{E}[\qz{Q}{h_1}^2 \qz{Q}{h_2,h_3}^2] \\
&= 3 \lambda_{h_1}^2 + \lambda_{h_1}
 + \lambda_{h_1}\big(\lambda_{h_1,h_2}(1-\delta(h_1-h_2)) \\
&\quad + \lambda_{h_2,h_3}(1-\delta(h_2-h_3))\big).
\end{aligned}
\end{equation}
where 
\begin{equation*}
    \mathrm{E}[\qz{Q}{h_1}^2 \qz{Q}{h_1,h_2}^2] = \lambda_{h_1} 3 \lambda_{h_1,h_2}(1-\delta(h_1-h_2)),
\end{equation*}
\begin{equation*}
    \mathrm{E}[\qz{Q}{h_1}^2 \qz{Q}{h_2,h_3}^2] = \lambda_{h_1} 3 \lambda_{h_2,h_3}(1-\delta(h_2-h_3)),
\end{equation*}
and 
\begin{equation*}
    \mathrm{E} [\qz{Q}{h_1}^4] = 3 \lambda_{h_1}^2 + \lambda_{h_1}.
\end{equation*}

\bibliographystyle{aasjournal}
\bibliography{main}

\end{document}